\documentclass[a4paper,10pt]{article}
\usepackage{amsmath}
 \usepackage{amssymb}
\usepackage{verbatim}
\usepackage{float}
\usepackage[small]{caption}
\usepackage[a4paper, left=2.5cm, right=2.5cm, top=3.5cm, bottom=3.5cm, headsep=1.2cm]{geometry}

\newcommand{\be}{\begin{equation}}
\newcommand{\ee}{\end{equation}}
\newcommand{\bea}{\begin{eqnarray}}
\newcommand{\eea}{\end{eqnarray}}
\newcommand{\bal}{\begin{align}}

\newcommand{\enl}{\end{align}}

\def\ov{\over  }

\newif\ifpdf
	\ifx\pdfoutput\undefined
	\pdffalse % we are not running PDFLaTeX
	\else
	\pdfoutput=1 % we are running PDFLaTeX
	\pdftrue
	\fi

	\ifpdf
	\usepackage[pdftex]{graphicx}
         \else
	\usepackage{graphicx}
	\fi

\begin{document}

{\bf{Lower bound for the ratio  of charged wino tracks to  charged lepton tracks at LHC.}}

\vskip0.5cm

Giorgio Calucci  (Physics Department of the Trieste University  \footnote{ email: giorgio@ts.infn.it }) and Roberto Iengo  (SISSA  and INFN, Trieste \footnote{ email: iengo@sissa.it })

\vskip1cm

{\bf{Abstract}}: By using kinematical arguments we derive a theorem for the ratio: [production cross-section of charged-neutral wino pairs plus anything else] over [production cross-section of charged-neutral lepton pairs plus anything else].  
We do that by working out the consequences of substituting the lepton pair with a wino pair,
leaving untouched everything else of the interaction.
We apply this result to the possible production of winos at LHC obtaining a  lower bound for the ratio of the number of charged-winos over charged-leptons tracks and also find the average charged-wino track length, within a region in the relevant parameter space (e.g. available energy and mass). 

\vskip1cm

{\bf{General framework.}}

\vskip0.5cm

One relevant hypothesis is that the dark matter particles are non-relativistic neutral winos, namely the neutral member of a triplet of (very) massive fermions in the adjoint representation of the weak 
$SU(2)_W$, 
and which can only interact with the $SU(2)_W$ vector bosons. This picture is per se worthwhile, even independently of the assumption of supersymmetry.

This hypothesis makes the dark matter very dark, because there is no tree-level elastic  scattering of neutral winos on ordinary matter,
because they are not coupled to the neutral vector boson, and the non relativistic inelastic scattering on ordinary matter giving neutral-wino $\to$ charged-wino is not allowed  
%for charged winos even a little heavier than the neutral winos.
even for a tiny charged-neutral mass difference of the order of tens of $MeV$. 
 Therefore, the interaction of dark matter with ordinary matter
can only occur at the order one-loop in the weak interactions.

This is a somewhat extreme hypothesis, but one can also relax it imagining that the neutralino is a linear combination of a wino and a higgsino (that is a member 
of the fundamental representation of $SU(2)_W$). This would make dark matter less dark as tree level elastic scattering is possible in this case. Actually,
a study of the dark-matter relic density indicates that in the $TeV$ mass-range dark matter should be predominantly pure wino \cite{ullioiengo}.

We consider here the possible production of a pair made of a charged and a neutral wino, plus anything else. This process comes from the coupling of the pair to
a virtual $W^\pm$ vector boson  (this is so even in the case of a mixing with higgsino, because its coupling to higgs is proportional to the weak coupling constant
and we neglect the higgs coupling to the ordinary matter). 

The key point is that the production of a charged(electron or muon)-neutral(neutrino) lepton (one of them being anti-lepton) pair, plus anything else,  also  comes  from the coupling of the pair to
a virtual  $W^\pm$ vector boson  (neglecting the coupling to higgs).
Therefore, one can work out the consequences of substituting that lepton pair with the wino pair of the same total energy
\footnote{we neglect radiative corrections of the pair final state},
leaving  everything else of the  to be the same. In this way one avoids to make any assumption on the dynamics of the production 
of the vector boson and of any other accompanying  processes. 

To summarize, we do not attempt to find the cross-section for the production of winos , but only to find a lower bound for the ratio of the production rate of  wino pairs to the rate of   lepton pairs. 
In other words, our aim is to state that, if there are winos of a given mass $m_w$,  the observation of a production  rate 
of seemingly unpaired charged leptons with momentum $q$ (the neutral lepton of the pair being unobserved)
implies the observation of a rate of charged wino tracks larger than a bound depending on $m_w/q$. We have in mind the direct observation of those tracks, besides the observation of the products of the decay charged-wino $\to$ neutral wino, and we also derive bounds for the lengths of these charged wino tracks.

In particular, we can consider the possible production of winos at LHC. 
If the winos are quite massive the total energy of the pair will be quite high and since we compare their rate of production with the rate of charged-neutral lepton pairs of the same total energy, these lepton pairs 
will presumably come from a W-boson directly produced in the parton interaction (a process usually called DY), rather than from the secondary decay of other produced particles. 
However, even if that picture is rather plausible, we are not obliged to make such an assumption. Our results depend on the ratio of the wino mass to the lepton energy, and when this ratio is larger than 1 
there is simply no possibility of wino production.

\vskip0.5cm

{\bf{Main results.}}

\vskip0.5cm

We  consider the relation among the production  of a charged-neutral lepton-antilepton pair (say $l^\pm l^0=$ either $e^{-}\bar\nu_e$, or $ e^+\nu_e$, or $ \mu^{-}\bar\nu_\mu$, or  $\mu^+\nu_\mu$) and the production of a charged wino-neutral wino pair 
(either $w_{ino}^-w_{ino}^0$, or  $w_{ino}^+w_{ino}^0$). 
We  take the center-of-mass frame of the  pair and  indicate 
%by $\pm\vec q_{CM}$  the three-momentum of the leptons, by $\pm\vec q_{wCM}$ he three-momentum of the winos (in the approximation of equal mass), and 
by $Q_0$ the total (lepton or wino) pair energy in this frame. 

%In the case of the lepton pair, in the massless approximation, $q_{CM}=Q_0/2$ whereas for the wino pair $q_{wCM}=\sqrt{Q_0^2/4-m_w^2}$, 
%the wino mass being $m_w$ in the approximation of the same mass for the charged and neutral one.

Our main result, derived in the Appendix A,  is that 
\be
R\equiv{\sigma_{w_{ino}^\pm w_{ino}^0}\ov\sigma_{l^\pm l^0}}~\geq \hat R\equiv  4\sqrt{1-{4 m_w^2\ov Q_0^2}}
\ee
where $\sigma_{w_{ino}^\pm w_{ino}^0}$ and $\sigma_{l^\pm l^0}$ are, respectively, the cross section 
\footnote{ we write $\sigma$ for short, in the place of the more precise ${d\sigma\ov dQ_0^2}$, both for the winos and the leptons.}
for producing a charged-wino neutral-wino pair, plus anything else, and for producing a charged-lepton neutral-lepton pair, plus anything else, with the same squared invariant mass of the pair $=Q_0^2$.
 Here, we have assumed the approximation of considering massless leptons and taking the same mass $m_w$ for the charged and neutral wino.

In the possible case of dark matter being a mixing of wino and higgsino of the same mass $m_w$: 
$\cos(\alpha)|wino>+\sin(\alpha)|higgsino>$,
the only difference with the previous
result is that the factor $4$ should be replaced by $4 \cos(\alpha)^2 +2 \sin(\alpha)^2$.

The ratio $\hat R$ gives a lower bound for the expected number of charged wino tracks, once a number of charged lepton tracks has been observed.
\vskip0.3cm
This result  is expressed in terms of quantities defined in the CM frame of the lepton pair, whereas the observations are made in the CM frame of the incident particles whose collision gives rise to the process in question, for instance the proton-proton CM frame in the case of LHC.
One would then need to perform a Lorentz transformation parametrized by $\vec\beta$, the relative velocity of the two fames. In order to know $\vec\beta$ one should reconstruct the momentum of the neutrino partner of the observed charged lepton. 

Nevertheless, we will see that one can obtain quite useful and stringent bounds on the possible wino observation even in the total ignorance of $\vec\beta$. Of course, by inserting further assumptions on the range of $\vec\beta$, motivated e.g. in the case of LHC by parton distributions and the like, those bounds tighten, however they do not significantly alter the information.
In order to show that, one has to work out the kinematics relating the CM frame of the lepton pair and the CM frame of the incident particles.

By doing that, our result can be expressed in terms of the ratio $x\equiv {m_w^2\ov q^2}$, where $ q$ is the observed charged-lepton momentum in the CM frame of the incident particles, and 
of the so far unknown $\beta$ and $z\equiv {\vec\beta\cdot\vec q\ov \beta q}$.  From the computations of the Appendix B we get:
\be
R~\geq \hat R\equiv  4\sqrt{1-{4 m^2_{w}\ov Q^2_0}} =4\sqrt{1-{m^2_{w}\ov q^2}{1-\beta^2\ov  (1-\beta z)^2}}
\ee
Really, if ${4 m^2_{w}\ov Q^2_0}>1$ there is no possible production of winos. Therefore a precise statement is 
\be
R~\geq \hat R(x,\beta,z)\equiv 4RealPart\sqrt{1-x{1-\beta^2\ov (1-\beta z)^2}},~~~with~~~x\equiv { m^2_{w}\ov q^2}
\ee
Assuming the forward -backward and cylindrical symmetry of the collision, $\vec\beta$ and $-\vec\beta$, and thus $z$ and $-z$,  are equally likely.  Therefore, the average value of $\hat R$ for
a given $|z|$ is
\be
<\hat R>(x,\beta,|z|)= \Big( \hat R(x,\beta,z)+ \hat R(x,\beta,-z) \Big)/2
\ee

%\vskip0.5cm

In conclusion, from the observation of a number $N_l(q)$ of tracks of charged leptons with a momentum $q$ measured in the CM frame of the collider (the neutral member of the pair being unobserved), we predict the minimal amount of charged wino tracks to be 
$\hat R\times N_l(q)$. In the following, by letting $\vec\beta$ vary in its range, we get, for a given $m_w/q$,  a region of values for $<\hat R>$.

\vskip0.5cm

Another interesting quantity is the expected length of the charged wino track. 
The dominant mode is the decay of the $\pm$charged wino into a $\pi^\pm$ and the neutral wino and the lifetime of a charged wino at rest is \cite{ibe} 
\be
\tau_0= {m_\pi m_\mu^2 (1-{m_\mu^2\ov m_\pi^2})^2\ov \Gamma(\pi^\pm\to\mu^\pm \nu_\mu) 
16 \delta m^3 (1-{m_\pi^2\ov \delta m^2})^{1/2}   } \label{life}
\ee
(where $ \delta m$ is the charged-neutral wino mass difference). 

Therefore the mean track length of the charged wino is
\be
L=\tau_0{\beta_w\ov\sqrt{1-\beta_w^2}} \label{length}
\ee
where $\beta_w$ is the velocity of the charged wino in the CM frame of the incident particles: \\
${\beta_w\ov\sqrt{1-\beta_w^2}}=\sqrt{{q_{0w}^2\ov m_w^2}-1}$,
$q_{0w}$ being its energy in this frame.
Also in this case, we need the information on  $\vec\beta$ to compute $q_{0w}$.

The charged-neutral wino mass difference has been computed to be $ \delta m=160 MeV$ in refs.\cite{arcadiullio} \cite{randallstrassler} for wino mass of the order of $TeV$, and somewhat higher say $162 MeV$ including two loops radiative corrections in ref. \cite{ibe}, giving $c \tau_0= 6.4~ cm$ from eq.(\ref{life}). 
 From eq.(\ref{length}) we get
the mean track length (in units of centimeters)
\be
L\big(x,\beta,z) =6.4 \sqrt{{q^2_{0w}\ov m^2_w}(x,\beta,z)-1} 
\ee
 where  by the kinematical computations of Appendix B one can write  ${m^2_{w}\ov q^2_{0w}}$  as a function of $x,\beta,z$.
 
For a given $|z|$ we expect to see on average a tract length 
\be 
<L>(x,\beta,|z|)= { \hat R(x,\beta,z)L(x,\beta,z)+ \hat R(x,\beta,-z)L(x,\beta,-z)\ov
\hat R(x,\beta,z)+ \hat R(x,\beta,-z) }
\ee

\vskip1cm

In the figures 1) and 2) we show our results for $<\hat R>$ and $<L>$ for the pure wino case, for $m_w^2/q^2=0.5$ fig.1, and for $m_w^2/q^2=0.8$ fig.2.

In the left part of the figure 
we allow $\beta, |z|$ to vary in their full range $~0\leq \beta\leq 1,~0\leq |z|\leq 1$, for the sake of illustrating the variability of the results, even though $\beta\sim 1$ 
would not be reached at LHC energies for wino masses of the order of $TeV$.

In the right part of the figure we  restrict the range 
\footnote{in the Appendix C we discuss the possible ranges of $\beta,z$.}
to $~~0\leq\beta\leq 0.4,~~0\leq |z|\leq \cos{\pi/6}~~$ (right part of the figure).

\begin{figure}[ht]\label{fig.1}
\centerline{\includegraphics[width=3.5in]{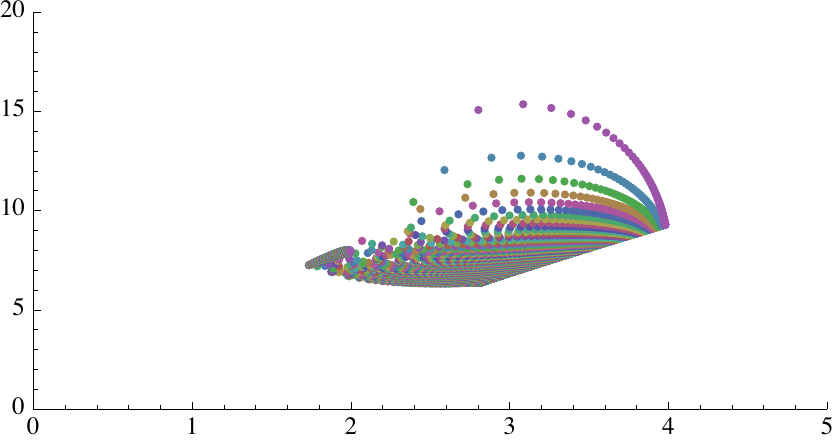} \ \ \ \  \includegraphics[width=3.5in]{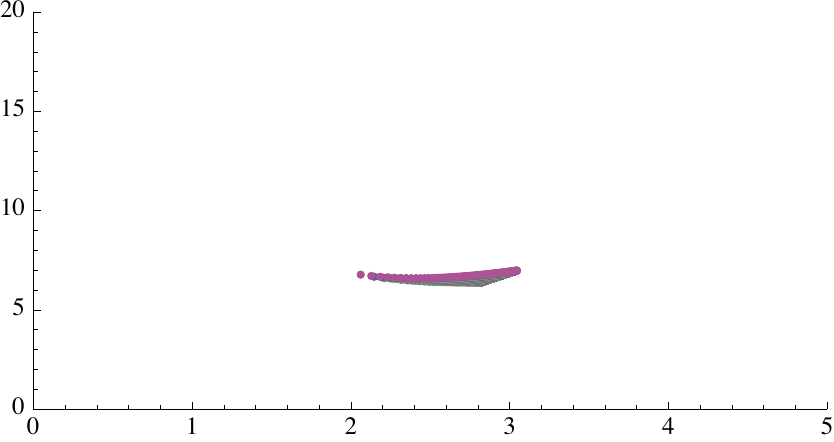} }
\caption{region of bounds for wino observation for $m_w^2/q^2=0.5$. 
\\ $~~~~~~~~~~~~~$ Left: full range for $\beta, |z|.~~$ Right: for $0\leq\beta\leq 0.4,~0\leq |z|\leq \cos{\pi/6}$. \\
$~~~~~~~~~~~~~$ x-axis $<\hat R>$ , y-axis $<L>$  (in centimeters).}
%{\centerline{\includegraphics[width=3in]{yukl0a10b10}}}
\end{figure}

%\newpage

\begin{figure}[ht]\label{fig.2}
\centerline{\includegraphics[width=3.5in]{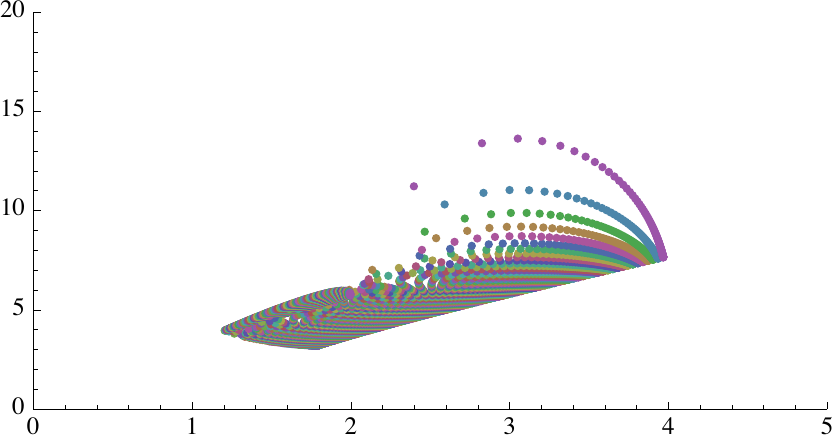} \ \ \ \  \includegraphics[width=3.5in]{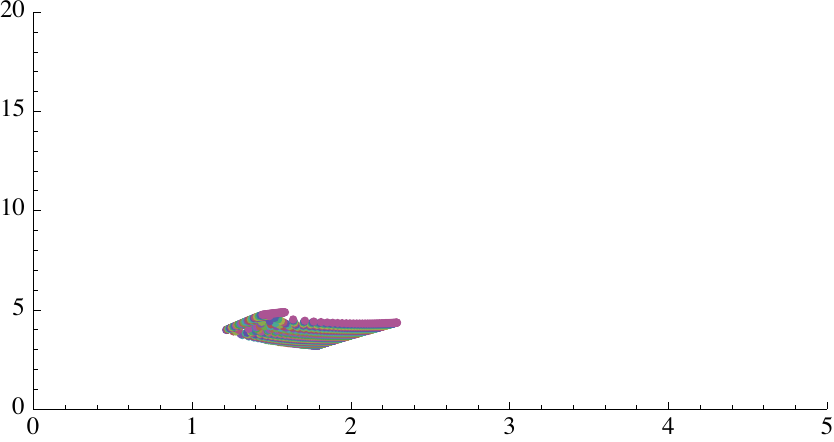} }
\caption{region of bounds for wino observation for $m_w^2/q^2=0.8$. 
\\ $~~~~~~~~~~~~~$ Left: full range for $\beta, |z|.~~$ Right: for $0\leq\beta\leq 0.4,~0\leq |z|\leq \cos{\pi/6}$. \\
$~~~~~~~~~~~~~$ x-axis $<\hat R>$ , y-axis $<L>$  (in centimeters).}
\end{figure}

\newpage

We can also compute the deviation of the expected track of the wino with respect to the direction of the observed track of the lepton.
Namely, in terms of the differential cross-sections at a given polar angle in the LHC frame, the statement is:
\be
d\sigma_{wino}(\theta'_{LHC})\geq 4 \sqrt{1-{m^2_{w}\ov q^2}{1-\beta^2\ov  (1-\beta z)^2}}~ d\sigma_{lepton}(\theta_{LHC})
\ee
We want to get informations on the possible angular difference $\delta\theta_{LHC}=\theta'_{LHC}-\theta_{LHC}$ that is the angle difference of the directions of $\vec q_w$ and $\vec q$.
We can estimate this difference by referring the polar angles to the direction of $\vec\beta$.
As we have seen, in the $CM$ frame we take the same emission direction for the lepton and the wino, therefore $z_{CM}$ is the same for both.
But due to the mass difference of the lepton and the wino, the emission direction is different in the LHC frame, therefore $z_w$, defined by $\vec q_w\cdot\vec\beta=q_w\beta z_w$,
is different from $z$, and we can work out its expression $z_w(x,\beta,z)$, see the Appendix B.

We can then compute the angle difference $\delta\theta_{LHC}$ of the direction of $\vec q_w$ and $\vec q$: 
\be
|\delta\theta_{LHC}|_z=| \arccos[z]-\arccos[z_w(x,\beta,z)]|
%\arccos[z_w(x,\beta,z) z+\sqrt{1-z_w(x,\beta,z)^2}\sqrt{1-z^2}]
\ee

Since, as we said, he sign of $z$ is not defined, we take here too the average, like we did in defining $<L>$ :
\be
<|\delta\theta_{LHC}|>_{|z|}={ \hat R_z |\delta\theta_{LHC}|_z+ \hat R_{-z} |\delta\theta_{LHC}|_{-z}\ov \hat R_z+\hat R_{-z}}
\ee
Our results for $<|\delta\theta_{LHC}|>_{|z|}$ are reported in the following figures:

\begin{figure}[ht]\label{fig.3}
\centerline{\includegraphics[width=2.5in]{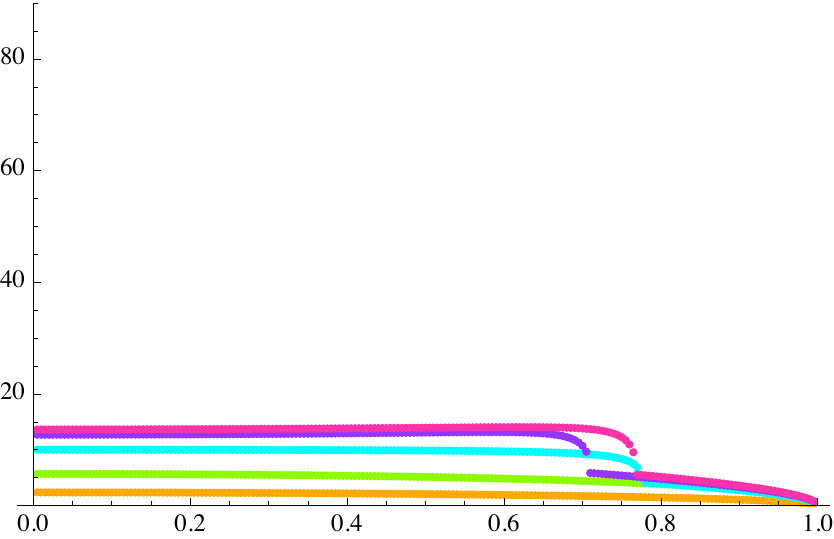} \ \ \ \  \includegraphics[width=2.5in]{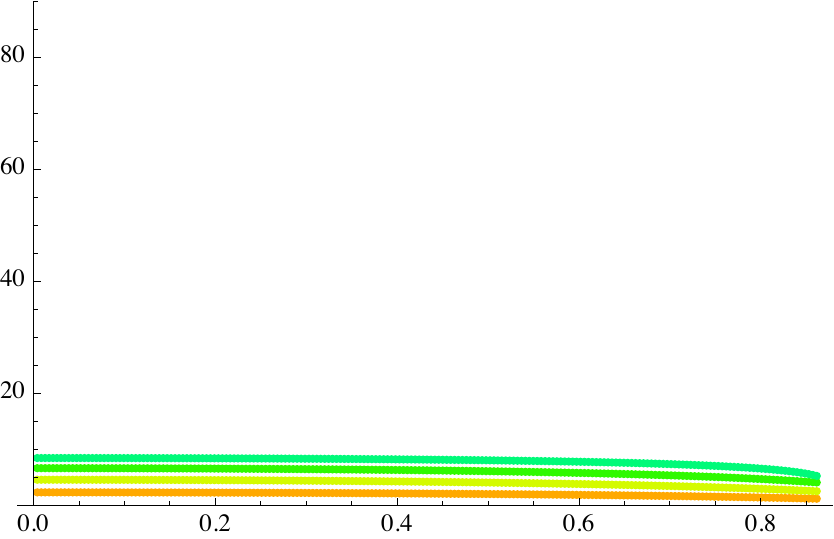} }
\caption{region of $\delta\theta_{LHC}$ for $m_w^2/q^2=0.5$ as a function of $|z|$ for various $\beta$ (different colors). 
 \\ $~~~~~~~~~~~~~$ Left: full range for $\beta, |z|.~~$ Right: for $0\leq\beta\leq 0.4,~0\leq |z|\leq \cos{\pi/6}$. \\
$~~~~~~~~~~~~~$ x-axis $|z|$, y-axis $<|\delta\theta_{LHC}=\theta_{wino}-\theta_{lepton}|>$ (in degrees). }
\end{figure}

\begin{figure}[ht]\label{fig.4}
\centerline{\includegraphics[width=2.5in]{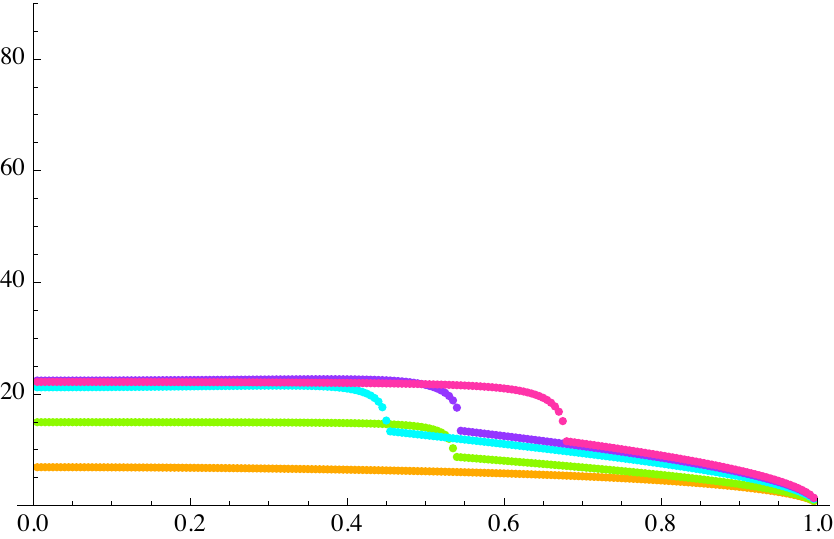} \ \ \ \  \includegraphics[width=2.5in]{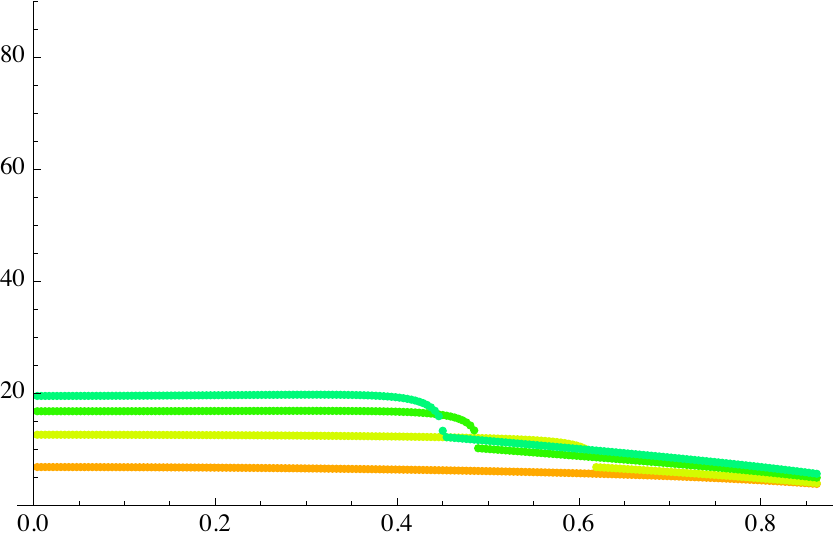}}
\caption{region of $\delta\theta_{LHC}$ for $m_w^2/q^2=0.8$ as a function of $|z|$  for various $\beta$ (different colors). 
 \\ $~~~~~~~~~~~~~$ Left: full range for $\beta, |z|.~~$ Right: for $0\leq\beta\leq 0.4,~0\leq |z|\leq \cos{\pi/6}$. \\
$~~~~~~~~~~~~~$ x-axis $|z|$, y-axis $<|\delta\theta_{LHC}=\theta_{wino}-\theta_{lepton}|>$ (in degrees). }
%{\centerline{\includegraphics[width=3in]{yukl0a10b10}}}
\end{figure}

\newpage

%\vskip0.5cm
{\bf{Conclusions.}}
\vskip0.5cm

Our conclusion is that if seemingly unpaired charged leptons are observed at LHC, the neutral lepton of the pair being unobserved,
(of course this would be a rare observation if their energy is very high, e.g. in the $TeV$ range), then also tracks of charged weakly interacting particles in the adjoint representation of $SU(2)_W$, with mass less than the lepton momentum, 
should be observed, assuming they exist and have a lifetime comparable to what has been predicted in the literature. 

We see from the figures that the number of charged wino tracks are expected to be from greater-or-equal-than  up to four-times-as-much-as the number of lepton tracks, and their length of the order of various centimeters, from 2.5 up to 7, and higher for larger $\beta$,  for wino-mass to lepton-momentum ratio in the range $0.7-0.9$ or less. Those charged wino tracks are expected to end giving rise to a charged lepton track, due to the weak decay charged wino $\to$ neutral wino.

Otherwise, if  no tracks of that kind are observed, our results can be used to put bounds on the existence of that kind of dark matter.

\vskip0.5cm

{\bf{Acknowledgments}}
We would like to thank Piero Ullio for having read our paper and for his useful comments.

\vskip0.5cm

{\bf{Appendix A: derivation.}}

\vskip0.3cm

Consider first the lepton pair $l^\pm l^0$.
Let us parametrize the momentum of the lepton in the lepton-pair $CM$ frame by
\be
q_{CM,\mu}=q_{CM} \big( 1,\sin(\theta)\cos(\phi),\sin(\theta)\sin(\phi),\cos(\theta) \big)
\ee
where $\theta,\phi$ are angles with respect to arbitrary axes. 
\vskip0.5cm
In this $CM$ frame the chiral spinor,anti-spinor $u^c,v^c$ representing the massless leptons are, in the representation in which $\gamma_5$ is diagonal:
$\gamma_5=Diag\{-1,-1,1,1\}$,
\bea
&u^c(\vec q_{CM})& = \sqrt{2 q_{CM}}(e^{-i\phi}sin(\theta/2),-cos(\theta/2),0,0) \\ \nonumber
&v^c(-\vec q_{CM})& = \sqrt{2 q_{CM}}(~~~~cos(\theta/2),e^{i\phi}sin(\theta/2),0,0)
\eea
and the chiral lepton-pair current in this frame is 
\bea
J_\mu^{lepton}(\theta,\phi) &=& \bar v^c(-\vec q_{CM})\gamma_\mu{1-\gamma_5\ov 2} u^c(\vec q_{CM}) \\ \nonumber
&=& 2 q_{CM} \big(0,\cos^2({\theta\ov 2})-e^{-2i\phi}\sin^2({\theta\ov 2}),-i(\cos^2({\theta\ov 2})+e^{-2i\phi}\sin^2({\theta\ov 2})),-e^{-i\phi}\sin(\theta) \big)
\eea

For reference, note also the anti-chiral spinor,anti-spinor $u^a,v^a$
\bea
&u^a(\vec q_{CM})& = \sqrt{2 q_{CM}}(0,0,\cos(\theta/2),e^{i\phi}\sin(\theta/2)) \\ \nonumber
&v^a(-\vec q_{CM})& = \sqrt{2 q_{CM}}(0,0,-e^{-i\phi}\sin(\theta/2),\cos(\theta/2))
\eea
and the corresponding anti-chiral current
\be
J^a_\mu(\theta,\phi)=\bar v^a(-\vec q_{CM})\gamma_\mu{1+\gamma_5\ov 2} u^a(\vec q_{CM})={J_\mu^{lepton}(\theta,\phi)}^*
%=-e^{-2 i\phi}J_\mu^{lepton}(\theta+\pi,\phi)
\ee

Consider now the  $w_{ino}^\pm w_{ino}^0$ pair (in the approximation in which the charged and the neutral wino have equal mass $m_w$) in the $CM$ frame of the pair with the same total energy $Q_0=2 q_{CM}$ 
as the leptons, taking the wino momentum $\vec q_{wCM}$ parallel to the lepton one $\vec q_{CM}$, with $q_{wCM}=\sqrt{ q^2_{CM}-m^2_w}$.
Their four-momentum will be:
\be
q_{wCM,\mu}=q_{wCM}\big( {q_{CM}\ov q_{wCM}}, \sin(\theta)\cos(\phi),\sin(\theta)\sin(\phi),\cos(\theta) \big)
\ee

One can choose the orthogonal basis of the helicity eigenstates for the spinors,anti-spinors representing the various possible states of the wino in this frame.
In this basis, there are states which we call pseudo-chiral ($pc$) :
\bea 
u^{pc}(\vec q_{wCM})= N_{u^{pc}}(e^{-i\phi}\sin{\theta\ov 2},-\cos{\theta\ov 2},{-q_{wCM}+q_{CM}\ov m_w}e^{-i\phi}\sin{\theta\ov 2},{q_{wCM}-q_{CM}\ov m_w} \cos{\theta\ov 2}) \\ \nonumber
v^{pc}(-\vec q_{wCM}) = N_{v^{pc}}(\cos{\theta\ov 2},e^{i\phi}\sin{\theta\ov 2},{q_{wCM}-q_{CM}\ov m_w}\cos{\theta\ov 2},{q_{wCM}-q_{CM}\ov m_w}e^{i\phi}\sin{\theta\ov 2}) 
\eea
and states which we call pseudo-antichiral ($pa$):
\bea
u^{pa}(\vec q_{wCM}) = N_{u^{pa}}(\cos{\theta\ov 2},e^{i\phi}\sin{\theta\ov 2},{q_{wCM}+q_{CM}\ov m_w}\cos{\theta\ov 2},{q_{wCM}+q_{CM}\ov m_w}e^{i\phi}\sin{\theta\ov 2}) \\ \nonumber
v^{pa}(-\vec q_{wCM})= N_{v^{pa}}(e^{-i\phi}\sin{\theta\ov 2},-\cos{\theta\ov 2},{-q_{wCM}-q_{CM}\ov m_w}e^{-i\phi}\sin{\theta\ov 2},{q_{wCM}+q_{CM}\ov m_w} \cos{\theta\ov 2})
\eea
with $N_{u^{pc}}^2=N_{v^{pc}}^2={ m_w^2\ov q_{CM}-q_{wCM}}$ and $N_{u^{pa}}^2=N_{v^{pa}}^2={ m_w^2\ov q_{CM}+q_{wCM}}$.
Note that $u^{pc,pa}\to u^{c,a}, v^{pc,pa}\to v^{c,a} $  in the limit $q_{CM}\to\infty$. 

\vskip0.5cm

In this basis we get four possible currents namely $J_\mu^{pc,pc},J_\mu^{pa,pa},J_\mu^{pc,pa},J_\mu^{pa,pc}$, where $J_\mu^{i,j}=\bar v^i \gamma_\mu u^j$
(the winos have pure vectorial coupling to the $SU(2)_W$ gauge bosons).

The key point of the derivation is that
\bea
J_\mu^{pc,pc}(\theta,\phi)=2 q_{CM} \big(0,\cos^2({\theta\ov 2})-e^{-2i\phi}\sin^2({\theta\ov 2}),-i(\cos^2({\theta\ov 2})+e^{-2i\phi}\sin^2({\theta\ov 2})),-e^{-i\phi}\sin(\theta) \big)=J_\mu^{lepton}(\theta,\phi) \\ \nonumber
J_\mu^{pa,pa}(\theta,\phi)=2  q_{CM} \big(0,\cos^2({\theta\ov 2})-e^{2i\phi}\sin^2({\theta\ov 2}),i(\cos^2({\theta\ov 2})+e^{2i\phi}\sin^2({\theta\ov 2})),-e^{i\phi}\sin(\theta) \big)={J_\mu^{lepton}(\theta,\phi)}^*
\eea

Since the cross-sectionis the sum of the square modulus of the amplitudes corresponding to the different currents, the cross-section for producing the wino pair will be larger
than what one gets by taking the charged and the neutral wino to be both $pc$ or both $pa$, that is by restricting the wino-currents to $J_\mu^{pc,pc},J_\mu^{pa,pa}$:
\be
\sigma_{wino}=\sigma(pc,pc)+\sigma(pa,pa)+\sigma(pc,pa)+\sigma(pa,pc)\geq
\sigma(pc,pc)+\sigma(pa,pa)
\ee
One can express the cross-section in a Lorentz invariant way by integrating the modulus square of a relativistic invariant amplitude over the relativistic invariant final particles phase space (and dividing by 
a Lorentz invariant factor representing the incident flux times the initial particles' density). 
Therefore, one can evaluate the  cross-section for producing the lepton pair by performing  the final phase space integration in the CM frame of the pair at a fixed $Q_0^2$ 
(here and in the following we write $\sigma$ for short in the place of $d\sigma/dQ_0^2$) and write
\be
\sigma_{lepton}=\int |K^\mu J_\mu^{lepton}(\theta,\phi)|^2 d\phi \sin\theta d\theta
%d\Omega 
\ee
whereas for the cross-section for producing the wino pair one has
\be
\sigma(pc,pc)=\sigma(pa,pa)=2 \int |K^\mu J_\mu^{lepton}(\theta,\phi) |^2 d\phi \sin\theta d\theta \sqrt{1-{4 m_w^2\ov Q_0^2}}
\ee
with the same $K^\mu $, whatever it is, the factor $2$ coming from the fact that the lepton vertex is ${g\ov\sqrt{2}}\bar v\gamma_\mu{1-\gamma_5\ov 2}u$ and the wino vertex is
$g\bar v\gamma_\mu u$. One can check by explicit integration that 
\be
\int_0^{2\pi} d\phi \int_0^\pi d\theta  \sin\theta |K^\mu J_\mu^{pc,pc}(\theta,\phi) |^2 = \int_0^{2\pi} d\phi \int_0^\pi d\theta  \sin\theta |K^\mu J_\mu^{pa,pa}(\theta,\phi) |^2 ={32\pi\ov 3}  q^2_{CM}( |K_1|^2+|K_2|^2+|K_3|^2)
\ee
from which the equality $\sigma(pc,pc)=\sigma(pa,pa)$ follows.
Therefore
\be
\sigma(pc,pc)+\sigma(pa,pa)=4\int |K^\mu J_\mu^{lepton}|^2 d\phi \sin\theta d\theta \sqrt{1-{4 m_w^2\ov Q_0^2}}=4\sigma_{lepton}\sqrt{1-{4 m_w^2\ov Q_0^2}}
\ee
and finally
\be
\sigma_{wino}\geq\sigma(pc,pc)+\sigma(pa,pa)=4 \sqrt{1-{4 m_w^2\ov Q_0^2}}~\sigma_{lepton}
\ee
Therefore
\be
R\equiv{\sigma_{wino}\ov\sigma_{lepton}}~\geq \hat R\equiv  4\sqrt{1-{4 m_w^2\ov Q_0^2}}
\ee
In the possible case of dark matter being a mixing of wino and higgsino of the same mass $m_w$: 
$\cos(\alpha)|wino>+\sin(\alpha)|higgsino>$,
the only difference with the previous
result is that the factor $4$ should be replaced by $4 \cos(\alpha)^2 +2 \sin(\alpha)^2$, since the higgsinos can be 
rearranged to form a Dirac doublet vectorially coupled to the $W-$boson with coupling $g/\sqrt{2}$ 
(we can neglect the higgs contribution, since the higgs boson coupling to the higgsino is also proportional to $g$ and the 
higgs coupling to the incoming matter matter is exceedingly small).

\vskip0.5cm

{\bf{Appendix B: kinematical computations.}}

\vskip0.3cm

We call $\vec q$  the three-momentum of the observed charged lepton and $q_0$ its energy in the CM frame of the incident particles. For the lepton, in the massless approximation,
$q_0=q$. We call $\vec q_w$ the wino three-momentum and $q_{0w}$ its energy in the CM frame of the incident particles , with $q_w=\sqrt{q_{0w}^2-m_w^2}$. 

By calling $\vec\beta$ the velocity of the $CM$ pair frame with respect to the  frame and defining $z$ by 
\be
\vec q\cdot\vec\beta=q\beta z
\ee
 we have
\be
q_{CM}= \gamma(q-\vec\beta\cdot\vec q)=q{1-\beta z\ov\sqrt{1-\beta^2}}
\ee
From that we get:
\be
{m^2_{w}\ov q^2_{CM}}={m^2_{w}\ov q^2}{1-\beta^2\ov  (1-\beta z)^2}
\ee

As for $\beta_w$ ( the velocity of the charged wino in the LHC frame to be used for determining the track length), it can be obtained from the Lorentz transformations:
\be
{ q_{0w}\ov m_{w}}={ {q_{CM}\ov m_{w}}+{\vec\beta\cdot\vec q_{CM}\ov q_{CM}}\sqrt{{q_{CM}^2\ov m_w^2}-1}\ov\sqrt{1-\beta^2}},
~~~{q_{CM}\ov m_{w}}={1\ov\sqrt{x}}{1-\beta z\ov\sqrt{1-\beta^2}}, ~~~{\vec\beta\cdot\vec q_{CM}\ov q_{CM}}=\beta {z-\beta \ov\ 1-\beta z} \label{kin}
\ee
(Note: $ q_{0w}\ov m_{w}$ is minimal for    ${\vec\beta\cdot\vec q_{CM}\ov q_{CM}}=-\beta$ and  ${q_{CM}\ov m_{w}}={1\ov\sqrt{1-\beta^2}}$, where it is $=1$ ).
  From eq.(\ref{kin}) we get ${q^2_{0w}\ov m^2_{w}}$  as a function of $x,\beta,z$: 

\be
{q^2_{0w}\ov m^2_{w}}(x,\beta,z)= {[(1-\beta z)^2-\beta(\beta-z)  \sqrt{ (1 - \beta z)^2-(1-\beta^2)x}~ ]^2  \ov (1-\beta^2)^2 (1-\beta z)^2 x}
\ee
 
 Next, by defining $z_{CM}$ by $ \vec q_{CM}\cdot\vec\beta=q_{CM}\beta z_{CM}$ we get from eq.(\ref{kin})
\be
z_{CM}={z-\beta\ov 1-z\beta}
\ee
As we have seen, in the $CM$ frame we take the same emission direction for the lepton and the wino, therefore $z_{CM}$ is the same for both.
But due to the mass difference of the lepton and the wino, the emission direction is different in the LHC frame, therefore $z_w$, defined by $\vec q_w\cdot\beta=q_w\beta z_w$,
is different from $z$.
In fact, by the Lorentz transformation 
\bea
\vec\beta\cdot\vec q_w=\beta q_w z_w={\vec\beta\cdot\vec q_{wCM}+\beta^2 q_{CM}\ov\sqrt{1-\beta^2}}={\beta q_{wCM}z_{CM}+\beta^2 q_{CM}\ov\sqrt{1-\beta^2}} \\ \nonumber
\to z_w(x,\beta,z)= ({\sqrt{q_{CM}^2/m_w^2-1}\ov \sqrt{q_{0w}^2/m_w^2-1}}z_{CM}+\beta {q_{CM}/m_w\ov\sqrt{q_{0w}^2/m_w^2-1}}) {1\ov\sqrt{1-\beta^2}}
\eea
and from that we can work out the expression $z_w(x,\beta,z)$.

\vskip0.5cm
{\bf{Appendix C: possible ranges for $\beta, z$.}}
\vskip0.3cm

We have seen that, even though $\beta, |z|$ are not known, we can find a region of values for $<\hat R>,<L>, <\delta\theta_{LCH}>$ for a given $x$, by varying $\beta,|z|$ in their full range:
$0\leq\beta\leq 1,~~0\leq |z|\leq 1$.
It can be nevertheless worthwhile to discuss which can be the most likely ranges for $\beta,|z|$. Here we take the case of LHC and we assume that the relevant process is DY.

Let us discuss the expected magnitude of $\beta$ and its direction. It can originate from 1)  mismatch of the almost opposite incoming parton momenta and/or 
2) the momentum of the produced vector-boson.
\vskip0.3cm
In the case 1) the direction of $\vec\beta$ is essentially the LHC beam direction. Let us call $\zeta=\hat s/s$ where $s$ is the nominal squared energy of the improved LHC, say $s=(14~ TeV)^2$, and $\hat s$ is the squared energy of the colliding partons.
We have $\zeta=x y$ where $x,y$ are the momentum fraction carried by the collidng partons, and the velocity of the colliding parton frame is
$\beta={x-y\ov x+y}={\zeta- y^2\ov \zeta+ y^2}$.
We are interested in the case where $\zeta$ is sizable, say $\zeta \geq 0.6$.  On average, we get (from the code \cite{code})  $\beta\sim 0.19$ for $\zeta=0.6$, $\beta\sim 0.08$ for $\zeta=0.8$. 
\vskip0.3cm

In the case 2) the maximum value of $\beta$ is reached when in the final state there is the virtual vector boson (producing the lepton or wino pair) and a massless object.
Actually, the dominant process is the collision of a gluon with a quark producing a vector-boson and a quark. By taking the quark as massless,
this process is enhanced by a would-be collinear divergence of the diagram having a pole in the Mandelstam $u$-variable
\footnote{The process gluon-gluon$\to$vector boson $via$ a quark loop would imply the same graph as the decay of a spin 1 particle into two massless vectors, which is forbidden (\cite{Yang}).},
when the vector-boson is anti-parallel to the direction of the incoming quark, that is the beam direction . 
This being so, in both cases 1) and 2) $\vec\beta$ can be taken along the beam direction.

As for its magnitude, $\beta={\hat s- Q_0^2\ov \hat s+Q_0^2}$. Remembering from Appendix B that $Q_0^2=4 q^2(1-\beta z)^2/(1-\beta^2)$ we get $\beta_{MAX}\sim 0.55$ for $4 q^2/\hat s=0.8$
and $\beta_{MAX}\sim 0.35$ or $4 q^2/\hat s=0.9$.

From those considerations, we have taken  $0\leq\beta\leq 0.4$ as a reasonable range.

As for $z$, that is the cosine  of the angle $\theta_{lepton}$ of the observed lepton with the direction of $\vec\beta$, assuming this direction is the one of the LHC beam, we may imagine to limit the observation to tracks which are outside a cone around the beam axis and limit that angle to be $-\pi/6\leq \theta_{lepton}\leq \pi/6$.

\end{document}